\documentstyle[aps,prl,twocolumn]{revtex}
\input epsf

\begin{document}
\draft
\title{The experimental observation of Beliaev damping in a Bose condensed gas}
\author{E. Hodby, O.M. Marag\`o, G. Hechenblaikner and C.J. Foot}
\address{Clarendon Laboratory, Department of Physics, University of Oxford,\\
Parks Road, Oxford, OX1 3PU, \\
United Kingdom.}
\date{\today}

\maketitle

\begin{abstract}
We report the first experimental observation of Beliaev damping of
a collective excitation in a Bose-condensed gas. Beliaev damping
is not predicted by the Gross-Pitaevskii equation and so this is
one of the few experiments that tests BEC theory beyond the mean
field approximation. Measurements of the amplitude of a high
frequency scissors mode, show that the Beliaev process transfers
energy to a lower lying mode and then back and forth between these
modes. These characteristics are quite distinct from those of
Landau damping, which leads to a monotonic decrease in amplitude.
To enhance the Beliaev process we adjusted the geometry of the
magnetic trapping potential to give a frequency ratio of 2 to 1
between two of the scissors modes of the condensate. The ratios of
the trap oscillation frequencies $\omega_y / \omega_x$ and
$\omega_z / \omega_x$  were changed independently, so that we
could investigate the resonant coupling over a range of
conditions.
\end{abstract}


In the Beliaev damping process \cite{Beliaev}, one quantum of
excitation converts into two quanta of a lower frequency mode, in
a manner analogous to parametric down-conversion of light in
nonlinear media. This process occurs readily in a homogeneous
system such as superfluid helium where there are many very closely
spaced low-lying energy levels. However, it has not previously
been observed for a trapped Bose-condensed gas, which has discrete
modes with well-resolved excitation energies. We have observed the
Beliaev process for a scissors mode \cite{GOS} of a BEC of
rubidium atoms, when the mode resonantly couples to another mode
at half its frequency. Beliaev damping can be regarded as a four
wave mixing process, in which a Bogoliubov quasiparticle interacts
with the ground state component of the condensate to produce two
new quasiparticles that share the initial energy equally.

At zero temperature, only Beliaev damping may occur, whilst at
finite temperature Landau damping is the primary mechanism which
dissipates the energy of collective excitation into the thermal
cloud \cite{Morgan}. Landau damping corresponds to a scattering
process between two quasiparticles in the initial mode, which
results in one particle gaining all of the energy and the other
going into the condensate ground state. The Landau damping rate
and its strong dependence on temperature has been measured for two
of the quadrupole excitations of the condensate \cite{Jin2}. We
observe a comparable Landau damping rate for the scissors mode in
most trap geometries, except those where the frequencies of the
two scissors modes were matched and the Beliaev process was
dominant.

An atom in a harmonic trap has three dipole modes of oscillation
along each of the principle axes, corresponding to the three trap
frequencies $\omega_x$, $\omega_y$ and $\omega_z$. The centre of
mass motion of an interacting cloud of atoms, such as the Bose
condensate and any surrounding thermal cloud, has the same
eigenfrequencies by Kohn's theorem \cite{Kohn}. However the
frequencies of other collective excitations of the condensate,
including the quadrupole modes, do not correspond to those of a
harmonic oscillator because of the strong interatomic
interactions. Six of the low lying collective excitations are
described by the linearised set of equations:

\begin{equation}
\mbox{\boldmath $\ddot{q}$} = \mbox{\boldmath $S \: q$}
\end{equation}
where the six components of the vector {\boldmath{$q$}},

\begin{equation}
\mbox{\boldmath $q$} = (Q_{ii},Q_{jj},Q_{kk},Q_{ij},Q_{jk},Q_{ik})
\end{equation}

\noindent are elements of the quadrupole tensor, {\boldmath{$Q$}}.

\begin{equation}
Q_{ij}= \langle x_i x_j \rangle
\end{equation}

\begin{figure}
\begin{center}\mbox{ \epsfxsize 3in\epsfbox{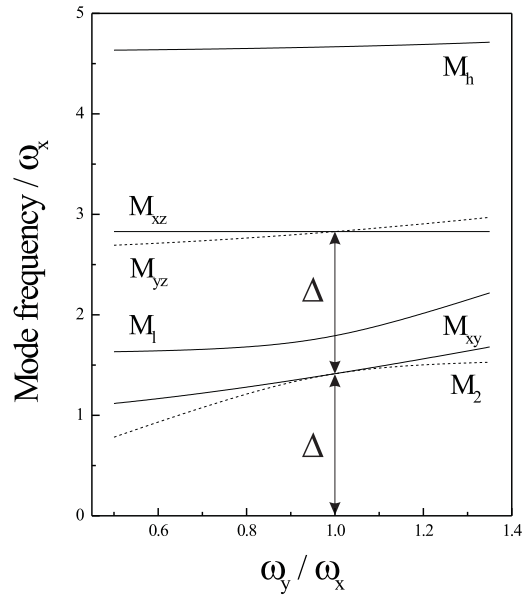}}\end{center}
\caption{The frequency spectrum of the six quadrupole modes of the
condensate as a function of the ellipticity of the trapping
potential in the xy plane. $\omega_z / \omega_x$ is chosen to be
$\sqrt{7}$, at which value $M_{xz}$ has exactly twice the
frequency of $M_{xy}$ and $M_2$, in an axially symmetric trap.
}\label{spec}
\end {figure}

 The six eigenvalues of {\boldmath{$S$}} give the frequencies of the six normal
 modes. Linear combinations of the diagonal elements of {\boldmath{$Q$}}
( e.g. $\langle x^2 \rangle$) describe the three normal modes with
principal axes of fixed orientation. We will refer to these modes
as $M_h, M_l, M_2$. In the case of the axially symmetric trap,
where the angular momentum about the axis (m) is a good quantum
number, these become the high and low lying $m=0$ modes and the
$m=2$ mode respectively \cite{Jin,Mewes}. The three independent
off-diagonal elements of {\boldmath{$Q$}} (e.g. $\langle x y
\rangle$) relate to the three scissors modes, which we label
$M_{xy}, M_{yz}, M_{xz}$, with frequencies $\omega_{xy},
\omega_{yz}, \omega_{xz}$. The frequency spectrum has been
calculated in \cite{Dalfovo,Pires} and is shown, in the
hydrodynamic limit, in Fig. \ref{spec} as a function of trap
geometry.

In general, the first and second harmonic oscillations involved in
the down-conversion process are the scissors modes in the xy and
xz planes respectively ($M_{xy}$ and $M_{xz}$). However, for the
special case of an axially symmetric trap, $M_{xy}$ is degenerate
with $M_2$ and the two cannot be distinguished. The frequencies of
these two scissors modes in the hydrodynamic limit are given by
$\omega_{xy} = \sqrt{\omega_x^2 + \omega_y^2}$ and $\omega_{xz} =
\sqrt{\omega_x^2 + \omega_z^2}$. Thus we observe a resonant
coupling when the following condition is satisfied:

\begin{equation}
2 \times \sqrt{1 + \frac{\omega_y^2}{\omega_x^2}} = \sqrt{1 +
\frac{\omega_z^2}{\omega_x^2}} \label{res}
\end{equation}

The scissors mode is a small angle, irrotational oscillation of
the condensate, about a given axis, that occurs without a change
of the cloud shape \cite{GOS,Marago}. $M_2$ is an out of phase
oscillation in the x and y directions, with a very small motion in
the z direction. This axial motion falls to zero amplitude in the
cylindrically symmetric trap. References \cite{Jin,Marago} show
the geometries of the two oscillations \cite{note}.

We initially create Bose-Einstein condensates in an axially
symmetric time-orbiting potential (TOP) trap ($\omega_x = \omega_y
= 127.6 \pm 1.0$ Hz). The trap frequency ratio, $\omega_z /
\omega_x$, is 2.83 and hence the condensate has an oblate shape.
After RF evaporative cooling, the condensate temperature is
significantly below 0.5 $T_c$ and the thermal cloud is no longer
visible. A detailed description of our method for selectively
exciting the xz scissors mode is given
 in \cite{Marago}. In summary, we apply an additional bias field of amplitude $B_z$ in the axial
direction, oscillating in phase with the x component of the TOP
bias field, $B_x$. This enables the symmetry axis of the trapping
potential to be tilted by an angle that depends on $B_z / B_x$. We
first form the condensate in a standard TOP trap, then
adiabatically tilt the trap to an angle $\phi$ and then suddenly
flip it to $-\phi$. This excites a scissors mode oscillation of
amplitude $ \theta = 2 \phi$ in the xz plane, about the new
equilibrium position.

\begin{figure}
\begin{center}\mbox{ \epsfxsize 3in\epsfbox{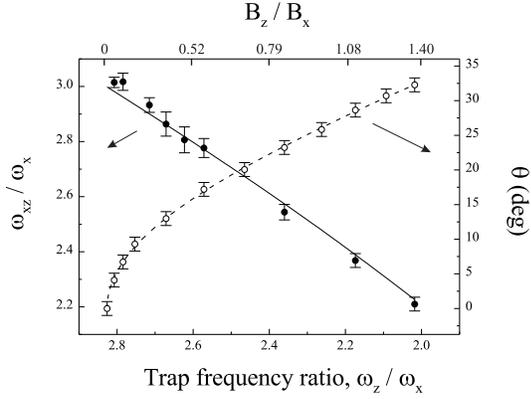}}\end{center}
\caption{The various effects of increasing the z bias field,
$B_z$. $B_z / B_x$ is shown along the top axis and the trap
frequency ratio, $\omega_z / \omega_x$, that results is shown
along the lower x axis. The solid line is the theoretical value of
the xz scissors frequency, $\omega_{xz}$, scaled against
$\omega_x$. The solid circles show experimental values obtained by
fitting sine waves to data such as in fig. \ref{damp}a,b (left
axis). The dotted line shows the theoretical excitation angle,
$\theta$, whilst the open circles show the values obtained from
measuring the tilt angle of the static trap (right
axis).}\label{freqang}
\end {figure}

However the application of the z bias field has a second, more
subtle effect \cite{Hodby}. The axial trapping frequency,
$\omega_z$, and hence the xz scissors mode frequency is reduced,
as $B_z$ is increased.The radial frequencies $\omega_x$ and
$\omega_y$ are also reduced, but by a negligable amount. By
changing $B_z / B_x$ we are able to tune the xz scissors frequency
into resonance with twice the xy scissors frequency. Since the
excitation angle is coupled to $B_z / B_x$, this is also changed,
but always remains well within the small angle scissors mode limit
defined in \cite{GOS}. Figure \ref{freqang} shows theoretically
and experimentally how the excitation amplitude, $\theta$, and the
frequency of the xz scissors mode, $\omega_{xz}$, change as a
function of $B_z / B_x$ and hence $\omega_z / \omega_x$.

Our first observation of Beliaev damping occurred in an axially
symmetric trap, $\omega_x = \omega_y$. In this case the xy
scissors mode becomes degenerate with, and indistinguishable from
the $\left| m=2 \right|$ quadrupole mode, with frequency $\sqrt{2}
\; \omega_x$ (Fig. \ref{spec}). The resonance condition of Eq.
(\ref{res}) is then fulfilled when $\omega_z / \omega_x =
\sqrt{7}$.

\begin{figure}
\begin{center}\mbox{ \epsfxsize 3in\epsfbox{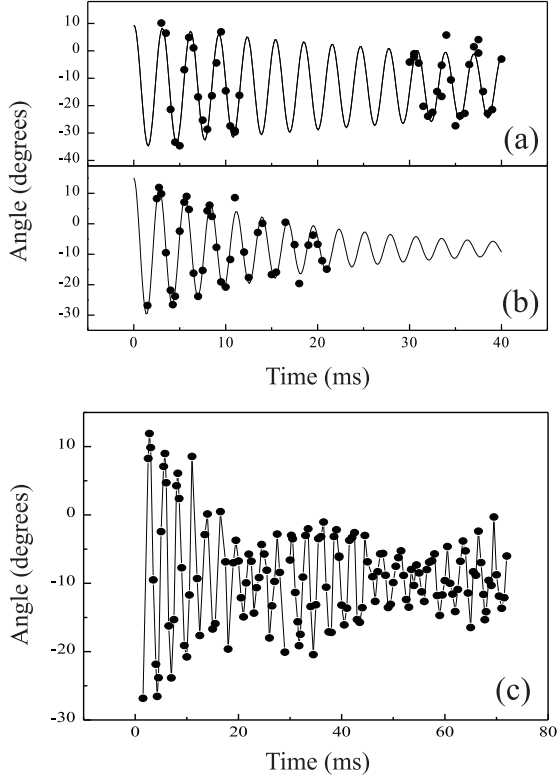}}\end{center}
\caption{Experimental data showing the angle of a condensate in
the xz scissors mode as a function of time. (a) and (b) are
typical plots from cylindrically symmetric traps with different
aspect ratios, fitted with exponentially decaying sine waves, from
which the damping rates, $\Gamma$, were extracted.\\ (a) $\omega_z
/ \omega_x = 2.36$, $\Gamma = 17 \pm 5$ Hz.
\\(b) $\omega_z / \omega_x = 2.62$,
$\Gamma = 57 \pm 8$ Hz.\\ In (b) the trap geometry is very close
to the resonance condition of Eq. \ref{res}. and the fast decay is
primarily due to Beliaev damping. The conditions of (a) are far
from resonance and so only the slow underlying Landau damping rate
is observed.
\\ Figure (c) has the same initial data as (b) but continues to
track the oscillation over a longer period. The transfer of energy
back and forth between the two modes is clearly observed, as the
amplitude of $M_{xz}$ falls and rises. }\label{damp}
\end {figure}

We recorded the xz scissors mode oscillation using destructive
absorption imaging. The angle of the cloud as a function of
oscillation evolution time, was extracted from 2D gaussian fits to
absorption images of the expanded condensate. We fit an
exponentially decaying sine wave to the angle versus time data and
use the damping rate as a measure of the rate at which energy is
flowing out of the mode. For values of $\omega_z / \omega_x$ far
from the resonance condition, we observe only a slow Landau
damping rate of $\sim$18 Hz (Fig. \ref{damp}a). This background
damping rate increases with the mode frequency, $\omega_{xz}$, as
predicted in \cite{Pit}.

\begin{figure}
\begin{center}\mbox{ \epsfxsize 3in\epsfbox{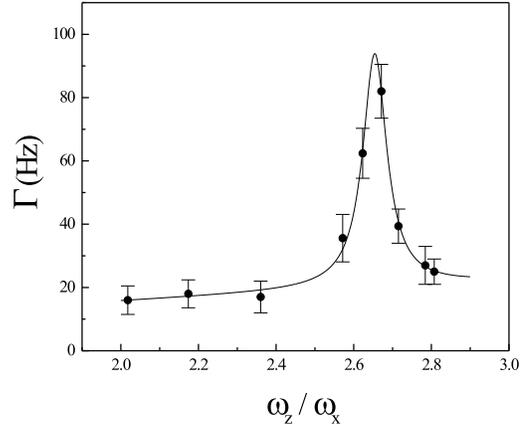}}\end{center}
\caption{Damping rate of the xz scissors mode as a function of the
trap frequency ratio $\omega_z / \omega_x$, in an axially
symmetric trap. The data is fitted with a Lorentzian peak plus a
linear function. The Lorentzian gives the trap geometry for peak
Beliaev damping. Far from resonance, the Landau damping rate is
observed. The increase of the Landau damping rate with mode
frequency is modeled by the linear part of the
fit.}\label{damppeak}
\end {figure}

Close to resonance we observe two new features in the xz scissors
mode. Firstly there is a sharp increase in the initial damping
rate of the oscillation, as the Beliaev process becomes
significant (Fig. \ref{damp}b). Secondly, over longer observation
periods ( $t
> 30$ms ), the amplitude envelope grows and then falls again, as
energy continues to be transferred back and forward between the
two modes, in a manner characteristic of non-linear coupling (Fig.
\ref{damp}c) \cite{Morgan2}. To obtain a measure of the total
damping rate, we fit the decaying sine wave up to, but not beyond,
the first amplitude minimum. The maximum coupling rate occurs when
$\omega_z / \omega_x = 2.65 \pm 0.04$, in very good agreement with
the predicted value of $\sqrt{7}$ (Fig. \ref{damppeak}).

In the axially symmetric trap described so far, $M_2$ and $M_{xy}$
are degenerate, but in a totally anisotropic trapping potential,
$\omega_x \neq \omega_y$, this degeneracy is broken (Fig.
\ref{spec}). To investigate coupling to the pure xy scissors mode,
we repeated the experiment, looking for resonant behaviour in a
range of triaxial traps. The trap geometry was chosen to be close
to the resonance condition between $M_{xy}$ and $M_{xz}$ given in
Eq. (\ref{res}).

\begin{figure}
\begin{center}\mbox{ \epsfxsize 3in\epsfbox{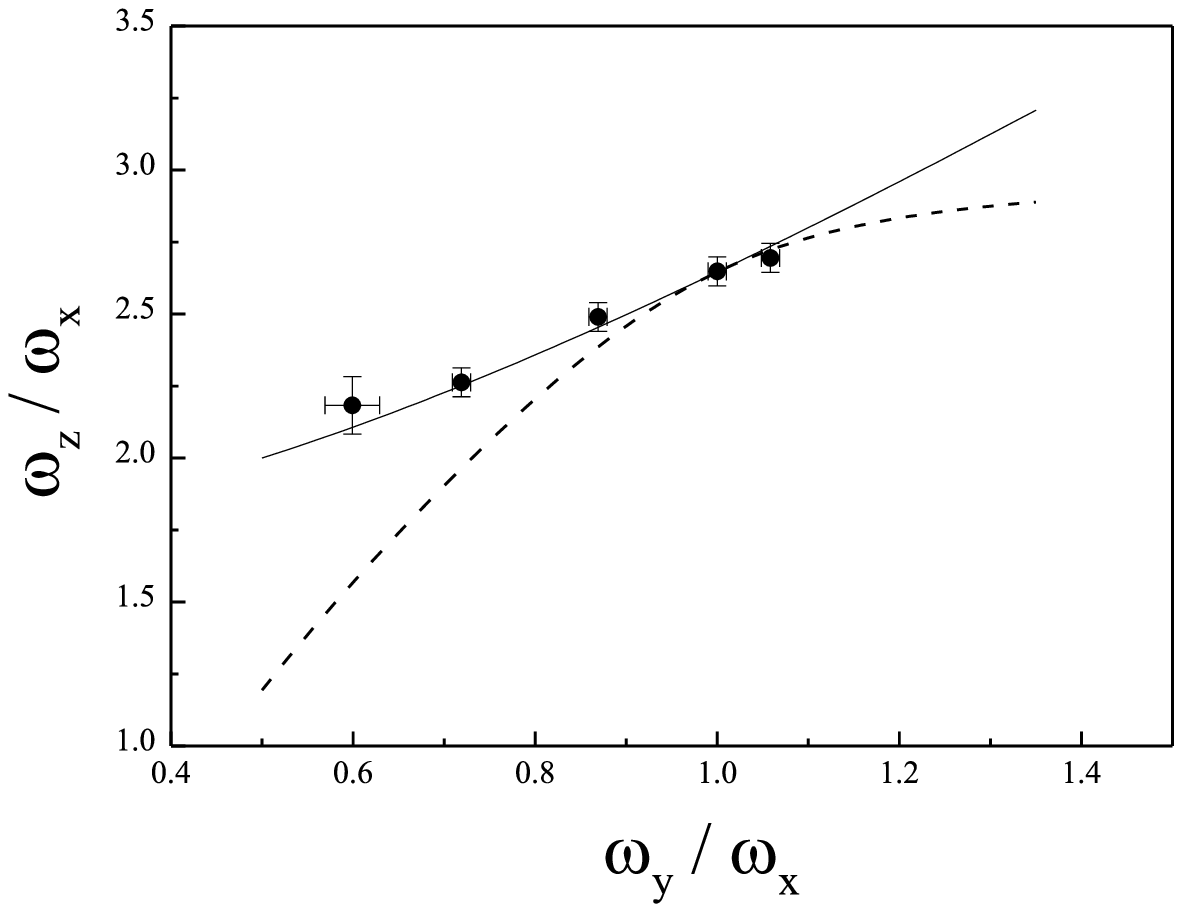}}\end{center}
\caption{Trap geometries for which $M_{xz}$ is resonant with twice
the frequency of $M_{xy}$ (solid line) and $M_{2}$ (dashed line).
The data points show the value of $\omega_z / \omega_x $ at which
the peak damping rate was observed, for a given $\omega_y /
\omega_x $.}\label{coupcurve}
\end {figure}

 The condensate was formed in a symmetric TOP trap as before.
However, whilst it was slowly tilted in the xz plane, it was also
adiabatically deformed by ramping the amplitude of the y component
of the TOP bias field. This made the trapping potential elliptical
in the xy plane, with $\omega_y / \omega_x$ between 0.6 and 1.05
\cite{Ensher}. For each value of $\omega_y / \omega_x$, we
determined the value of $\omega_z / \omega_x$ for peak coupling.
Figure \ref{coupcurve} shows this data, with the resonant coupling
condition superimposed. The agreement between experiment and
theory over a range of trap geometries confirms that the correct
damping process has been identified.

These observations do not rule out the possibility of
down-conversion into the $M_2$ mode in an anisotropic potential,
since our traps were specifically chosen for $M_{xz}$ to be
resonant with $M_{xy}$ and not $M_2$. In fact, for traps with a
very small anisotropy in the xy plane, so that {\it both} modes
are close to resonance with the xz scissors mode, we notice a
broadening of the peak in the initial damping rate. This is
expected when a second down-conversion process, into the $M_2$
mode, is resonant in a slightly different trap geometry.

In non-linear optics, it is well-known that down-conversion
produces output beams of squeezed light and analogous effects
 with matter waves may be explored using BEC. In previous work we observed
up-conversion between the two $m = 0$ modes of a BEC in a
cylindrically symmetric potential \cite{Hechenblaikner} and
interesting comparisons may be made between the two processes.
Whilst the up-conversion process showed strong dispersion of the
driving oscillation close to resonance, this was not observed in
the measured values of $\omega_{xz}$ in Fig. \ref{freqang}.
Secondly, the return of energy to the original mode, shown
strikingly in Fig. \ref{damp}c, was not observed for the
up-conversion process. Finally, whilst spontaneous up-conversion
is predicted by the Gross-Pitaevskii equation, the reverse process
of spontaneous down-conversion may not be described within this
framework. This may be explained by the following simple argument
\cite{Morgan2}. The GPE
 contains a nonlinear term proportional to $\left| \psi \right|^2$, so excitation
of a pure mode with time dependence $e^{i \omega t}$, leads to
terms of angular frequency $2 \omega$, but not to terms in $\omega
/ 2$ and hence no down-conversion. However, if there is initially
some amplitude at frequency $\omega / 2$, then nonlinear mixing
with the oscillation at frequency $\omega$ leads to a transfer of
energy between the modes at $\omega$ and $\omega / 2$.

In conclusion, we have observed Beliaev damping of the xz scissors
mode. Measurements over a range of trap geometries confirm that
this corresponds to resonant down-conversion into the xy scissors
mode. In future experiments, we intend to probe the nature of the
final squeezed state and investigate the role of the thermal cloud
in the down-conversion process at finite temperature.

We would like to thank all the members of the Oxford theoretical
BEC group, in particular K. Burnett, M. Rusch and S. Morgan for
their help and advice.

This work was supported by the EPSRC and the TMR program (No. ERB
FMRX-CT96-0002). O.M. Marag\`{o} acknowledges the support of a
Marie Curie Fellowship, TMR program (No. ERB FMBI-CT98-3077).

\begin{references}

\bibitem{Beliaev} S.T.Beliaev, Sov. Phys. JETP. {\bf 34}, 323
(1958)

\bibitem{BEC} M.H. Anderson {\it et al.}, Science {\bf 269}, 198
(1995); K.B. Davis {\it et al.}, Phys. Rev. Lett. {\bf 75}, 3969
(1995).

\bibitem{GOS} D. Gu\'ery-Odelin and S. Stringari, Phys. Rev. Lett.
{\bf 83}, 4452 (1999)

\bibitem{Morgan} S.A. Morgan J.Phys.B, {\bf 33} 3847 (2000)

\bibitem{Jin2} D.S. Jin, M.R. Matthews, J.R. Ensher, C.E. Wiemann,
E.A. Cornell, Phys. Rev. Lett. {\bf78}, 764 (1997);
D.M.
Stamper-Kurn, H-J. Meisner, S. Inouye, M.R. Andrews, W. Ketterle,
Phys. Rev. Lett. {\bf 81}, 500 (1998)


\bibitem{Kohn} W. Kohn, Phys. Rev. {\bf 123} 1242 (1961)

\bibitem{Jin}
          D.S. Jin, J.R. Ensher, M.R. Matthews, C.E. Wieman and E. A. Cornell,
          Phys. Rev. Lett. {\bf 77}, 420 (1996);
\bibitem{Mewes}
          M.O. Mewes, M.R. Andrews, N.J. van Druten, D.M. Kurn, C.G. Townsend and
          W. Ketterle, Phys. Rev. Lett. {\bf 77}, 988 (1996)

\bibitem{Dalfovo} F. Dalfovo, S. Giorgini, L.P. Pitaevskii
and S. Stringari, Rev. Mod. Phys. {\bf 71} 463 (1999)

\bibitem{Pires} M.O. da C. Pires and E.J.V. de Passos, J.Phys.B.
{\bf 33} 3929 (2000)

\bibitem{Marago} O.M. Marag\`{o}, S.A. Hopkins, J. Arlt, E. Hodby, G. Hechenblaikner and C.J. Foot, Phys. Rev. Lett. {\bf 84}, 2056 (2000)

\bibitem{note} Note that in an axially symmetric trap
down-conversion occurs into an equal superposition of the $m=+2$
and $m=-2$ modes, ensuring that the net angular momentum about the
z axis is zero at all times. For a triaxial trap, angular momentum
is not conserved and need not be considered in the coupling
process.

\bibitem{Hodby} E. Hodby, G. Hechenblaikner, O.M. Marag\`{o}, J.
Arlt, S. Hopkins and C.J. Foot, J.Phys.B. {\bf 33}, 4087 (2000)

\bibitem{Pit}
          L.P. Pitaevskii and S. Stringari, Phys. Lett. A {\bf 235}, 398 (1997);
          P.O. Fedichev, G.V. Shlapnykov, and J.T.M. Walraven,
          Phys. Rev. Lett. {\bf 80}, 2269 (1998)

\bibitem{Morgan2} S.A. Morgan, S. Choi, K. Burnett and M. Edwards,
Phys. Rev. A, {\bf 57}, 3818 (1998)

\bibitem{Ensher} J.R. Ensher, PhD thesis, University of Colorado
(1998);
 J. Arlt, O. Marag\`{o} , E. Hodby, S.A. Hopkins, G. Hechenblaikner, S. Webster
          and C.J. Foot, J. Phys. B {\bf 32}, 5861 (1999)

\bibitem{Hechenblaikner} G. Hechenblaikner, O.M. Marag\`{o}, E.
Hodby, J. Arlt, S. Hopkins and C.J. Foot, Phys. Rev. Lett. {\bf
85}, 692 (2000)

\end{references}
\end{document}